\documentclass[11pt,fleqn]{article}

\usepackage{a4}
\usepackage{amstext}
\usepackage{amsfonts}
\usepackage{amssymb}
\usepackage{bm}
\usepackage{color}
\usepackage{cite}
\usepackage{epsfig}
\usepackage{mathtools}
\usepackage{ulem}
\usepackage{caption}         
\usepackage{subcaption}   
\usepackage{framed}
\usepackage{multirow}
\usepackage{braket}
\usepackage{placeins}
\usepackage{soul}
\usepackage{comment}
\usepackage{dsfont}
\usepackage{hyperref}

\newcommand{\gtapprox}{\raisebox{-0.5ex}{$\,\stackrel{>}{\scriptstyle\sim}\,$}}

\begin{document}


\begin{flushright} TUM-EFT 144/21 \end{flushright}

\vspace{-0.2cm}

\begin{center}

{\huge \bf Lattice gauge theory computation of the}

{\huge \bf static force}

\vspace{0.5cm}

\textbf{Nora Brambilla$^a$, Viljami Leino$^a$, Owe Philipsen$^b$, Christian Reisinger$^{b,c}$, \\ Antonio Vairo$^a$, Marc Wagner$^{b,c}$}

$^a$~Technische Universit\"at M\"unchen, Physik Department, James-Franck-Stra{\ss}e 1, 85748 Garching, Germany \\
$^b$~Goethe-Universit\"at Frankfurt am Main, Institut f\"ur Theoretische Physik, Max-von-Laue-Stra{\ss}e 1, D-60438 Frankfurt am Main, Germany \\
$^c$~Helmholtz Research Academy Hesse for FAIR, Campus Riedberg, Max-von-Laue-Stra{\ss}e 12, D-60438 Frankfurt am Main, Germany

\vspace{0.5cm}

\end{center}

\begin{tabular*}{16cm}{l@{\extracolsep{\fill}}r} \hline \end{tabular*}

\vspace{-0.5cm}
\begin{center} \textbf{Abstract} \end{center}
\vspace{-0.4cm}

We explore a novel approach to compute the force between a static quark and a static antiquark with lattice gauge theory directly. 
The approach is based on expectation values of Wilson loops or Polyakov loops 
with chromoelectric field insertions. We discuss theoretical and technical aspects in detail, 
in particular,
how to perform their finite multiplicative renormalization and their
evaluation using a multilevel algorithm. 
We also compare the numerical results for the static force to the corresponding results obtained in the traditional way, i.e., by computing first the static potential and then taking the derivative.

\begin{tabular*}{16cm}{l@{\extracolsep{\fill}}r} \hline \end{tabular*}

\thispagestyle{empty}


\newpage

\setcounter{page}{1}

\section{Introduction}

The static quark-antiquark potential $V(r)$ is one of the best studied quantities in QCD. Very early after QCD was established, 
it was related to the large time behavior of Wilson loop expectation 
values~\cite{Wilson:1974sk,Susskind:1976pi,Fischler:1977yf,Brown:1979ya}. 
Wilson loop expectation values can be computed in lattice QCD. Indeed, they were among the first quantities studied numerically and can nowadays be computed very precisely. 
At small quark-antiquark separations $r$ the static potential, also referred to as QCD static energy, can also be calculated in a weak coupling expansion. 
The perturbative expression of the static energy is known quite accurately. 
The three loop contributions have been computed in Refs.\ \cite{Anzai:2009tm,Smirnov:2009fh}. 
At three loops the static energy contains also a term proportional to $\ln \alpha_{\rm s}$. 
This term has been computed in Refs.\ \cite{Brambilla:1999qa,Brambilla:1999xf} and resummed to all orders at leading logarithmic accuracy in Ref.\ \cite{Pineda:2000gza}. 
Next-to-leading logarithms have been computed and resummed to all orders in Refs.\ \cite{Brambilla:2006wp,Brambilla:2009bi}. 
At present, the static energy is, therefore, known at N$^3$LL (next-to-next-to-next-to-leading logarithmic) accuracy.

Combining the high precision results for $V(r)$ from lattice QCD as well as from perturbative QCD allows for an accurate extraction of the strong coupling $\alpha_{\rm s}$, 
that is competitive with lattice determinations using other observables \cite{Aoki:2019cca}. 
Recent extractions of $\alpha_{\rm s}$ from $V(r)$ can be found in Refs.\ \cite{Karbstein:2014bsa,Bazavov:2014soa,Karbstein:2018mzo,Takaura:2018vcy,Bazavov:2019qoo,Ayala:2020odx}.

The perturbative expression for the static potential computed in dimensional regularization is affected by a renormalon ambiguity of order $\Lambda_\text{QCD}$ \cite{Pineda:1998id,Hoang:1998nz}. 
In a lattice regularization, there is no renormalon ambiguity, but a linear divergence due to the self-energy, which is of order $\alpha_{\rm s}(1/a)/a$, where $a$ denotes the lattice spacing. 
The self energy vanishes order by order in dimensional regularization. 
Both the divergent self energy in a lattice regularization and the renormalon in dimensional regularization can be absorbed into an additive $r$-independent constant. 
Indeed, the physical information is contained in the shape of $V(r)$, which, after charge renormalization, 
is finite and renormalon free. The shape of the potential is the static force, defined via $F(r) = \partial_r V(r)$. The static force also carries the relevant information to extract $\alpha_{\rm s}$.

A possibility to compute the static force with lattice gauge theory is to first compute the static potential and then to take the derivative via finite differences. 
This program has been successfully carried out in quenched lattice QCD \cite{Necco:2001xg,Necco:2001gh}. 
In full QCD and at small separations, lattice data points for $V(r)$ are typically sparse and exhibit large discretization errors. 
One can still determine the static force by interpolating the lattice data points with a smooth function, but the interpolation might become a sizable source of uncertainty \cite{Bazavov:2014soa}.

In Refs.\ \cite{Vairo:2015vgb,Vairo:2016pxb} it has been recently suggested 
that the force between a static quark and a static antiquark can be computed directly from the expectation value of a Wilson loop with a chromoelectric field inserted in one of the temporal Wilson lines, 
a result originally derived in Ref.\ \cite{Brambilla:2000gk}. 
In this paper we carry out a quenched lattice QCD study of this quantity. 
We discuss technical aspects in detail, 
e.g.\ how to compute the static force using 
either Wilson or Polyakov loops with chromoelectric field insertions. 
We also compute the finite multiplicative renormalization of the insertions.
Moreover, we compare the efficiency of this novel method to compute the static force with the traditional method of first computing the static potential and then taking the derivative. 
Our conclusion is that the determination of the static force from the expectation value of a Wilson loop or a pair of Polyakov loops with a chromoelectric field insertion is a viable alternative method. 
Both methods provide consistent results with comparable errors, but different systematics. 
We note that the connection between the force and the color electric field has already been used e.g.\ in Refs.\ \cite{Baker:2018mhw,Baker:2019gsi} to determine the string tension in quenched lattice QCD.

The paper is organized in the following way. 
In section~\ref{sec:theory} we review the derivation of the force in terms of a Wilson loop with a chromoelectric field inserted in one of the temporal lines 
and we discuss the lattice discretization of this expression and its renormalization. In section~\ref{sec:lattice}, 
we specify our SU(3) lattice setup and we explain, how we use the multilevel algorithm \cite{Luscher:2001up} in our computations. 
The numerical results are presented in section~\ref{sec:result}. 
In particular, we compare the results obtained with our method to the corresponding results obtained by deriving the static potential and we discuss the efficiency of the two methods. 
In section~\ref{sec:conclusion} we draw some conclusions and give a brief outlook. 
Some details on the optimization of the parameters of the multilevel simulations can be found in Appendix~\ref{sec:appendix}. 
Preliminary results of this work have been presented at a recent conference~\cite{Brambilla:2019zqc}.


\section{\label{sec:theory}Theoretical and technical aspects}


\subsection{\label{sec:FE}The static force from the static potential}

The traditional way to compute the force between a static quark-antiquark pair is to take the numerical derivative
of the static potential,
\begin{eqnarray}
F(r) = \partial_r V(r) .
\end{eqnarray}
The latter is extracted from a non-perturbative evaluation of rectangular Wilson loops,
\begin{eqnarray}
\label{EQN764} {\rm Tr}\{{\rm P} \, W_{r \times T}\} = {\rm Tr}\bigg\{{\rm P} \, \exp\bigg(i g \oint_{r \times T} dx_\mu \, A_\mu(x)\bigg)\bigg\} ,
\end{eqnarray}
extending in space from $\bf{0}$ to $\mathbf{r}$ and in time from $-T/2$ to $+T/2$. 
They represent the correlation function of a static quark at $\bf{0}$ and an antiquark at $\mathbf{r}$ connected by a string of color flux. 
Here and in the following, time and fields are understood as Euclidean, ${\rm Tr}\{ \ldots \}$ denotes the color trace and ${\rm P}$ implies the path ordering prescription for the color matrices. 
The spectral decomposition of the Wilson loop expectation value reads
\begin{eqnarray}
\label{spcdecW} \langle {\rm Tr}\{{\rm P} \, W_{r \times T}\} \rangle = |a_0(r)|^2 e^{-E_0(r) T} + \sum_{n > 0} |a_n(r)|^2 e^{-E_n(r) T} .
\end{eqnarray}  
The coefficients $|a_n(r)|^2$ describe the overlap of the spatial Wilson lines with the eigenstates $| n_{\alpha \beta}(r) \rangle$ 
of the Hamiltonian of Yang--Mills theory in the presence of two static color charges in temporal gauge with energy $E_n$,
\begin{eqnarray}
H_0 | n_{\alpha \beta}(r) \rangle = E_n(r) | n_{\alpha \beta}(r) \rangle .
\end{eqnarray}
The color indices, $\alpha,\beta = 1,2,3$, indicate the transformation of the states in the fundamental and anti-fundamental representations at $\bf{0}$ and $\mathbf{r}$, respectively. 
The ground state energy is identified with the static potential, $V(r) \equiv E_0(r)$, and the excited states with $n > 0$ have energies $E_n(r) > V(r)$. 
Clearly, in the large $T$ limit, the terms with $n > 0$ in Eq.\ (\ref{spcdecW}) are exponentially suppressed with respect to the first one. 
For sufficiently large $T$, the Wilson loop expectation value thus allows us to extract $V(r)$ from the first term.

Instead of the Wilson loop, also a correlation function of Polyakov loops can be employed to compute the potential. 
Defining the Polyakov loop as a normalized trace of a temporal Wilson line closing through the periodic boundary,
\begin{eqnarray}
L({\bf x}) = \frac{1}{N_c} {\rm Tr} \bigg\{{\rm P} \, \exp\bigg(i g \int_0^T dt \, A_0(x)\bigg)\bigg\} ,
\end{eqnarray}
the expectation value of its correlation function has the spectral decomposition
\begin{eqnarray}
\label{EQN_LL_spectral_decomposition} \langle L^\dag({\bf 0}) L(\mathbf{r}) \rangle = \frac{1}{N_c^2} \sum_{n,\alpha,\beta} |\langle n_{\alpha \beta} | n_{\beta \alpha} \rangle|^2 e^{-E_n(r) T} = \frac{1}{N_c^2} \sum_n e^{-E_n(r) T}
\end{eqnarray}
(see, e.g., Ref.\ \cite{Jahn:2004qr}), 
again permitting us to extract $V(r)$ for large $T$ from the leading term.

\subsection{\label{sec:theory:analytic}The static force in terms of the chromoelectric field}
An alternative way to compute the static force was proposed in Refs.\ \cite{Vairo:2015vgb,Vairo:2016pxb} using the equation
\begin{eqnarray}
\label{EQN_F} F(r) = \partial_r V(r) = \lim_{T \rightarrow \infty} -i \frac{\langle {\rm Tr}\{{\rm P} \, W_{r \times T} \,\hat{\mathbf{r}} \cdot g \mathbf{E}(\mathbf{r},t^\ast)\}\rangle}{\langle \textrm{Tr}\{{\rm P} \, W_{r \times T}\} \rangle} .
\end{eqnarray}
Here $\hat{\mathbf{r}}$ is the spatial direction of the separation of the static color charges and $\mathbf{E}(\mathbf{r},t^\ast)$ 
denotes the chromoelectric field located on one of the temporal Wilson lines at a time $-T/2 < t^\ast < +T/2$. 
The chromoelectric field components are defined as $E_j(x) = F_{j 0}(x)$ in terms of the non-Abelian field strength tensor. 
In the limit $T \rightarrow \infty$, the right hand side of Eq.\ (\ref{EQN_F}) is independent of $t^\ast$, as long as $t^\ast$ is a fixed time.

The derivation of Eq.\ (\ref{EQN_F}) follows from Ref.\ \cite{Brambilla:2000gk}. First, we recall the identities
\begin{eqnarray}
\nonumber & & \hspace{-0.7cm} D_j(\mathbf{r},+T/2) \phi(\mathbf{r},+T/2;\mathbf{r},-T/2) = \\
\label{id1} & & = \phi(\mathbf{r},+T/2;\mathbf{r},-T/2) D_j(\mathbf{r},-T/2) + i \int_{-T/2}^{+T/2} dt \, \phi(\mathbf{r},+T/2;\mathbf{r},t) \, g E_j (\mathbf{r},t) \phi(\mathbf{r},t;\mathbf{r},-T/2) \\
\label{id2} & & \hspace{-0.7cm} D_i({\bf r},\pm T/2) \, \phi({\bf r},\pm T/2; {\bf 0},\pm T/2) = \phi({\bf r},\pm T/2;{\bf 0},\pm T/2) \partial_i + O_\pm                                                                                 
\end{eqnarray}
(see Refs.\ \cite{Eichten:1979pu,Brambilla:2000gk}), 
where $\phi(y;x)$ is a straight Wilson line connecting the point $x$ with the point $y$, $D_j(x) = \partial_j - i g A_j(x)$ 
is the gauge covariant derivative computed at the point $x$, $\partial_j = \partial / \partial x_j$ 
and $O_{\pm}$ are operators involving the chromomagnetic field on the Wilson lines at the times $\pm T/2$ (their explicit expression can be found in Ref.\ \cite{Brambilla:2000gk}). 
From this, one can derive
\begin{eqnarray}
\label{eq4}
\langle \partial_j {\rm Tr}\{{\rm P} \, W_{r \times T}\} \rangle \sim i \int_{-T/2}^{+T/2} dt \, \langle {\rm Tr}\{{\rm P} \, W_{r \times T} \, g E_j ({\bf r},t)\} \rangle ,
\end{eqnarray}
where $\sim$ denotes the asymptotical equivalence with respect to large $T$ (i.e. $f(T) \sim g(T)$ if and only if $\displaystyle \lim_{T \rightarrow \infty} f(T) / g(T) = 1$).
This is because the chromomagnetic fields in the spatial Wilson lines have no overlap with the lowest energy level in the spectrum of the static quark-antiquark pair, 
hence their contribution is exponentially suppressed in the large $T$ limit. Since in that limit we furthermore have $\langle {\rm Tr}\{{\rm P} \, W_{r \times T}\} \rangle \sim |a_0(r)|^2 e^{-V(r) T}$, 
we can conclude from Eq.\ (\ref{eq4}) that for large~$T$,
\begin{eqnarray}
\label{eq5} -\partial_j V(r) T  \sim  \frac{i}{\langle {\rm Tr}\{{\rm P} \, W_{r \times T}\} \rangle} \int_{-T/2}^{+T/2} dt \, \langle {\rm Tr}\{{\rm P} \, W_{r \times T} \, g E_j({\bf r},t)\} \rangle.
\end{eqnarray}  

The spectral decomposition of the expectation value of a Wilson loop with a chromoelectric field insertion reads~\cite{Brambilla:2000gk}
\begin{eqnarray}
  \nonumber & & \hspace{-0.7cm} \nonumber \langle {\rm Tr}\{{\rm P} \, W_{r \times T} \, g E_j({\bf r},t)\} \rangle =
                |a_0(r)|^2 e^{-V(r) T} \langle 0_{\alpha \beta} | g E_j^{\beta \gamma} | 0_{\gamma \alpha} \rangle \\
          & & \hspace{2cm} + \sum_{(n,m) \neq (0,0)} a_n^*(r) a_m(r) e^{-(E_n(r) + E_m(r)) T/2 + (E_n(r) - E_m(r)) t} \langle n_{\alpha \beta} | g E_j^{\beta \gamma} | m_{\gamma \alpha} \rangle ,
\label{spcdecWE_}              
\end{eqnarray}
where the greek indices label the fundamental representation color components and run from $1$ to $3$, cf.~section \ref{sec:FE}.
From the spectral decomposition it follows that
\begin{eqnarray}
 \int_{-T/2}^{+T/2} dt \, \langle {\rm Tr}\{{\rm P} \, W_{r \times T} \, g E_j ({\bf r},t)\} \rangle  &\sim&  T \,|a_0(r)|^2 e^{-V(r) T} \langle 0_{\alpha \beta} | g E_j^{\beta \gamma} | 0_{\gamma \alpha} \rangle 
                                                                                                             \nonumber\\
  &\sim& T \,\langle {\rm Tr}\{{\rm P} \, W_{r \times T} \, g E_j ({\bf r},t^*)\} \rangle \,.
\label{spcdecWE}  
\end{eqnarray}
The first line of Eq.~\eqref{spcdecWE} makes clear that the large $T$ limit of the left-hand side selects the ground state contribution.
The second line shows that evaluating the integral with the chromoelectric field at some fixed arbitrary time $-T/2 < t^* < T/2$ instead of $t$ 
is a valid leading order approximation in the large $T$ limit.
                     
Thus, in the large $T$ limit Eq.\ (\ref{eq5}) reduces to Eq.\ (\ref{EQN_F}),
\begin{eqnarray}
  \label{eq6} \partial_j V(r) \sim -i \frac{\langle {\rm Tr}\{{\rm P} \, W_{r \times T} \, g E_j ({\bf r},t^*)\} \rangle}{\langle {\rm Tr}\{{\rm P} \, W_{r \times T}\} \rangle}.
\end{eqnarray}
Equation \eqref{EQN_F} provides a way to compute the force directly from a Wilson loop expectation value
rather than to compute it by taking a numerical derivative of the static potential $V(r)$.
However, the spectral decomposition shows a faster convergence for the ordinary Wilson loop than for
$\langle {\rm Tr}\{{\rm P} \, W_{r \times T} \,\hat{\mathbf{r}} \cdot g \mathbf{E}(\mathbf{r},t^\ast)\}\rangle$.
In fact the exponential suppression of the sub-leading terms in  Eq.~\eqref{spcdecW}, which is $e^{-(E_n(r) - V(r)) T}$,
is stronger than the exponential suppression of the sub-leading terms in  Eq.~\eqref{spcdecWE}, which is $e^{-(E_n(r) - V(r)) T / 2}$.
Thus, it is not clear a priori, which observable is more efficient to evaluate numerically.

Just as in the case of the static potential, the static force may also be extracted from correlation functions of Polyakov loops instead of from Wilson loops. 
To this end we define a Polyakov loop with a chromoelectric field insertion,
\begin{eqnarray}
L_E(\mathbf{r}) = \frac{1}{N_c} {\rm Tr}\bigg\{{\rm P} \, \exp\bigg(i g \int_{t^*}^T dt \, A_0(x)\bigg) \hat{\mathbf{r}} \cdot g \mathbf{E}(\mathbf{r},t^*) {\rm P} \, \exp\bigg(i g \int_0^{t^*} dt \, A_0(x)\bigg)\bigg\} .
\end{eqnarray}
The analog of Eq.\ (\ref{EQN_F}) then reads
\begin{eqnarray}
\label{EQN_F_LL} F(r) = \lim_{T \rightarrow \infty} -i \frac{\langle L^\dag({\bf 0}) L_E(\mathbf{r}) \rangle}{\langle L^\dag({\bf 0}) L(\mathbf{r}) \rangle} .
\end{eqnarray}

The static force corresponds to a renormalization group invariant quantity, which does not require renormalization
and thus can be computed from bare fields. In order to see this, consider its calculation via the static potential 
extracted from ordinary, bare Wilson loops as described in section \ref{sec:FE}. 
The static potential is an eigenvalue of the Hamiltonian $H_0$ of Yang-Mills theory in the presence of static sources,
and as such contains an $r$-independent, linear divergence due to the self-energies of the static sources.
This divergence is removed by taking the $r$-derivative, so that the force,
$F(r) = \partial_r V(r)$, 
is the same when computed from either renormalized or unrenormalized Wilson loops.
Equation (\ref{EQN_F}) then implies the same to hold for the new observable on its right side. 


\subsection{\label{SEC_lattice}Lattice discretization}

The Wilson loop $W_{r \times T}$ is discretized by the standard product of adjacent link variables $U_\mu(x) = e^{i a g A_\mu(x)}$ 
around a rectangle with sides of length $r$ and $T$, where $r/a$ and $T/a$ are integers and $a$ denotes the lattice spacing. 
In the following we always choose separations $\mathbf{r}$ parallel to the $z$ axis. The static potential can then be obtained via
\begin{eqnarray}
\label{EQN_V_lat} V(r,a) = \lim_{T \rightarrow \infty}  V_\text{eff}(r,T,a) \quad , \quad V_\text{eff}(r,T,a) = -\frac{1}{a} \ln \frac{\langle \textrm{Tr}\{{\rm P} \, W_{r \times (T+a)}\}\rangle}{\langle \textrm{Tr}\{{\rm P} \, W_{r \times T}\}\rangle} .
\end{eqnarray}
$V(r,a)$ depends on $a$ and diverges in the continuum limit, because of the self-energy of the static quarks. One can get rid of this self-energy, 
by considering static potential differences, $V(r_1,a) - V(r_2,a)$, 
which have leading order discretization errors proportional to $a^2$, 
because the unimproved Wilson pure gauge action we are using in this work is accurate up to $O(a^2)$ \cite{Gattringer:2010zz}.

A straightforward way of computing the static force from the static potential is using a discretized derivative,
\begin{eqnarray}
\label{EQN_F_unimproved} F_{\partial V}(r,a) = \frac{V(r+a,a) - V(r-a,a)}{2 a} .
\end{eqnarray}
Discretization errors in static potential differences are known to be particularly large for small quark-antiquark separations $r$. 
A common procedure to reduce these is to define tree-level improved separations
\begin{eqnarray}
\label{EQN_r_I} r_I = r_I(r) = \bigg(\frac{2 a}{4 \pi (G(r+a)-G(r-a))}\bigg)^{1/2}
\end{eqnarray}
with
\begin{eqnarray}
G(r) = \frac{1}{a} \int_{-\pi}^{+\pi} \frac{d^3k}{(2 \pi)^3} \frac{\cos(r k_3/a)}{4 \sum_{j=1}^3 \sin^2(k_j/2)}
\end{eqnarray}
and to replace Eq.\ (\ref{EQN_F_unimproved}) by
\begin{eqnarray}
\label{eq:f2} F_{\partial V}(r_I,a) = \frac{V(r+a,a) - V(r-a,a)}{2 a}
\end{eqnarray}
(see Ref.\ \cite{Sommer:1993ce} for details). At tree-level one then obtains
\begin{eqnarray}
F_{\partial V}(r_I,a) = \frac{g^2}{4 \pi r_I^2} ,
\end{eqnarray}
in agreement with continuum perturbation theory.

The lattice formulation of the observable given by the right hand side of Eq.\ (\ref{EQN_F}) is straightforward. 
We place the field insertion $\mathbf{E}(\mathbf{r},t^*)$ in the numerator at $t^* = 0$, such as to maximize the distance to the temporal boundaries of the Wilson loop (which are located at $-T/2$ and $+T/2$). 
The discretization of the field insertion $E_j = F_{j 0}$ depends on the choice for the discrete partial derivative in the field strength tensor. 
For the simple forward derivative, $\partial_j f(x) = (f(x+a) - f(x)) / a$, the field strength tensor is related to the plaquette $P_{\mu,\nu}$ in the usual way,
\begin{eqnarray}
P_{\mu,\nu} = 1 + i a^2 g F_{\mu \nu} + \mathcal{O}(a^3) ,
\end{eqnarray}
which gives the chromoelectric lattice field components
\begin{eqnarray}
\label{eq:lat_Edef} g E_j = \frac{P_{j,0} - P^\dagger_{j,0}}{2 i a^2} + \mathcal{O}(a) .
\end{eqnarray}
A smaller discretization error can be achieved using the symmetric definition of the derivative, $\partial_j f(x) = (f(x+a) - f(x-a)) / 2 a$, and either a so-called butterfly
\begin{eqnarray}
\label{EQN_butterfly} \Pi_{j 0} = \frac{P_{j,0}+P_{0,-j}}{2}
\end{eqnarray}  
or a cloverleaf
\begin{eqnarray}
\label{EQN_cloverleaf} \Pi_{j 0} = \frac{P_{j,0} + P_{0,-j} + P_{-j,-0} + P_{-0,j}}{4}
\end{eqnarray}
of the plaquettes (see e.g.\ Ref.\ \cite{Bali:1997am}). For those cases the chromoelectric field is given by
\begin{eqnarray}
g E_j = \frac{\Pi_{j 0} - \Pi^\dagger_{j 0}}{2 i a^2} + \mathcal{O}(a^2) .
\end{eqnarray}
In our computations we use these symmetric discretizations (cloverleaf for Wilson loops, butterfly for Polyakov loops). With these definitions we arrive at the discretized version of Eq.\ (\ref{EQN_F}),
\begin{equation}
\label{eq:f1} F_E(r_I,a) = \lim_{T \rightarrow \infty} F_{E,\text{eff}}(r_I,T,a) \quad , \quad F_{E,\text{eff}}(r_I,T,a) = -i \frac{\langle \textrm{Tr}\{{\rm P} \, W_{r \times T} \, \hat{\mathbf{r}} \cdot g \mathbf{E}(\mathbf{r},t^*)\}\rangle}{\langle \textrm{Tr}\{{\rm P} \, W_{r\times T}\}\rangle}
\end{equation}
with $\mathbf{r}$ parallel to one of the spatial coordinate axes, 
and where we use again the tree-level improved separations $r_I$ defined in Eq.\ (\ref{EQN_r_I}) (one can show that at tree-level $F_E(r_I,a) = g^2 / 4 \pi r_I^2$, 
i.e.\ also in this case there is agreement with continuum perturbation theory). $F_E(r_I,a)$ depends on $a$ with leading order corrections proportional to $a^2$ 
(cf.\ similar observables discussed in Refs.\ \cite{Huntley:1986de,Bali:1997am,Bali:2000gf}).

With these choices, both ways to compute the static force formally have discretization errors of $O(a^2)$ and we expect
\begin{eqnarray}
F(r) = \lim_{a \rightarrow 0} F_{\partial V}(r_I,a) = \lim_{a \rightarrow 0} F_E(r_I,a) .
\end{eqnarray}
However, the approach to the continuum limit may well be 
quantitatively quite different for the two observables, because for operators involving elements of the field strength tensor, significant finite renormalization factors are expected 
when comparing the lattice regularization with regularization schemes in the continuum at values of the gauge coupling typically used in numerical simulations. 
The reason is the slow convergence of lattice perturbation theory, when expanded in the bare coupling~\cite{Lepage:1992xa}. 
This has been noted heuristically in early treatments of spin corrections to the static potential~\cite{Huntley:1986de,Bali:1997am,Koma:2006fw} 
and is observed both in a perturbative renormalization of the color-electric field correlator~\cite{Christensen:2016wdo} 
as well as in a non-perturbative renormalization of the color-magnetic field operator~\cite{Guazzini:2007bu}. 
Since Eq.\ (\ref{eq:f1}) involves a color-electric field, while the standard Wilson loop does not, 
we expect sizable differences between the two extractions of the static force at finite lattice spacing. 
To a large extent, these differences can, however, be absorbed into a multiplicative renormalization factor, 
\begin{eqnarray}
\label{EQN_ZE} Z_E(a) = \frac{F_{\partial V}(r_I^\ast,a)}{F_E(r_I^\ast,a)} ,
\end{eqnarray} 
where $r_I^\ast$ is an arbitrary separation, with $Z_E(a) \rightarrow 1$ for $a \rightarrow 0$. 
After determining this renormaliza\-tion factor at a single arbitrary separation $r_I^\ast$, it can be applied to $F_E(r_I,a)$ at all other separations,
\begin{eqnarray}
F_E^\text{ren}(r_I,a) = Z_E(a) F_E(r_I,a).
\end{eqnarray} 
$F_E^\text{ren}(r_I,a)$ should be significantly closer to both $F_{\partial V}(r_I,a)$ and $F(r)$ than is the case for $F_E(r_I,a)$, also at values of the gauge coupling typically used in numerical simulations.

All considerations of this section can be applied to correlation functions of Polyakov loops in an analogous way. Particularly important is the counterpart of Eq.\ (\ref{eq:f1}), which is
\begin{equation}
\label{EQN634} F_E(r_I,a) = \lim_{T \rightarrow \infty} -i \frac{\langle L^\dag(0) L_E(r) \rangle}{\langle L^\dag(0) L(r) \rangle} .
\end{equation}



\section{\label{sec:lattice}Lattice setup}


\subsection{Gauge link ensembles}

To discretize SU(3) Yang--Mills theory, we used the standard Wilson plaquette action. 
For the relation between the lattice spacing $a$ and the gauge coupling $\beta$ we took the parametrization from Ref.\ \cite{Necco:2001xg},
\begin{eqnarray}
\label{eq:sc} \ln(a/r_0) = -1.6804 - 1.7331 \, (\beta - 6) + 0.7849 \, (\beta - 6)^2 - 0.4428 \, (\beta - 6)^3 .
\end{eqnarray}

The parameters of the gauge link ensembles we generated for this work are collected in Table~\ref{TAB001}. 
We note that the spatial volumes $L^3$ are quite small with spatial extents $L \approx 1.2 \, \text{fm}$. 
Consequently, our results might be subject to sizable finite volume corrections. 
We do not consider this as a primary problem at this stage, 
because we were exploring and testing a new method, not trying to determine the static force or any other physical quantity precisely, i.e.\ for infinite volume.

To improve the ground state overlaps generated by the spatial Wilson lines in the Wilson loops, 
we used APE smeared spatial links with $\alpha_\textrm{APE} = 0.5$ and $N_\textrm{APE} = 50$ smearing steps (for detailed equations on APE smearing see e.g.\ Ref.\ \cite{APE:1987ehd,Jansen:2008si}).

\begin{table}[htb]
\begin{center}
\def\arraystretch{1.2}
\begin{tabular}{ccccc}
\hline
Ensemble & $\beta$ & $(L/a)^3 \times T/a$ & $r_0/a$ & $a$ $(\text{fm})$ \\
\hline
A & $6.284$ & $20^3\times 40$ & $\phantom{0}8.333$ & $0.060$ \\
B & $6.451$ & $26^3\times 50$ & $10.417$           & $0.048$ \\
C & $6.594$ & $30^3\times 60$ & $12.500$           & $0.040$ \\
\hline
\end{tabular}
\end{center}
\caption{\label{TAB001}Gauge link ensembles. To quote the lattice spacing in $\text{fm}$, we define $r_0 = 0.5 \, \text{fm}$.}
\end{table}


\subsection{\label{SEC005}Multilevel algorithm}

To compute in a very efficient way the correlation functions introduced in section~\ref{sec:theory}, 
i.e.\ Wilson loops and Polyakov loop correlation functions with and without chromoelectric field insertions, 
we used the multilevel algorithm \cite{Luscher:2001up}. 
We partitioned a lattice with $T/a$ lattice sites in temporal direction into $n_\text{ts}$ time-slices with thicknesses $p_1, p_2, \dots , p_{n_\textrm{ts}}$, where $\sum_j p_j = T/a$. 
For convenience we defined the time-slice partitioning $p_\textrm{ts} = \{p_1, p_2, \ldots, p_{n_p}\}$, 
which was repeated $N$ times to fill the lattice, i.e.\ $p_j = p_{j + n_p}$ and $n_\text{ts} = N n_p$. 
In principle, time-slices can be partitioned again, but throughout this work we used only a single level of partitioning.

Following a notation similar to that of Ref.\ \cite{Luscher:2001up}, 
correlation functions are written in terms of two-link operators $\mathds{T}(x,r \hat{j})_{\alpha \beta \gamma \delta} = \{U_0^*(x)\}_{\alpha\beta}\{U_0(x+r \hat{j})\}_{\gamma\delta}$, 
where $U_0(x)$ are link variables in the temporal direction and $\hat{j}$ denotes the unit vector in $j$-direction. 
In the context of the multilevel algorithm a regular Wilson loop is written as
\begin{eqnarray}
\label{eq:multilevel_Wilson} W_{r \times T}(x) = \mathds{L}(x,r\,\hat{j})_{\alpha \gamma} \{[\mathds{P}_1] [\mathds{P}_2] \dots [\mathds{P}_{n_\textrm{ts}}]\}_{\alpha \beta \gamma \delta} \mathds{L}^*(x+T \hat{0},r \hat{j})_{\beta \delta} ,
\end{eqnarray}
where $\mathds{L}(x,r \hat{j})$ is an APE smeared spatial Wilson line,
\begin{eqnarray}
\label{EQN633} \mathds{P}_k = \mathds{T}(x+(d_k-p_k) a \hat{0},r \hat{j}) \mathds{T}(x+(d_k-p_k+1) a \hat{0},r \hat{j}) \dots \mathds{T}(x+(d_k-1) a \hat{0},r \hat{j})
\end{eqnarray}
and $[\mathds{P}_k]$ is the average of $\mathds{P}_k$ in the time-slice extending from $d_k - p_k$ to $d_k$ with $d_k = \sum_{j=1}^k p_j$. 
In Eq.\ (\ref{eq:multilevel_Wilson}) and Eq.\ (\ref{EQN633}) we have used the multiplication prescription for two-link operators,
\begin{align}
\{\mathds{T}_1 \mathds{T}_2\}_{\alpha \beta \gamma \delta} = \{\mathds{T}_1\}_{\alpha \sigma \gamma \rho} \{\mathds{T}_2\}_{\sigma \beta \rho \delta} .
\end{align}
As discussed below in more detail, $[\mathds{P}_k]$ corresponds to an average over $n_m$ sublattice configurations separated by $n_u$ heatbath sweeps, 
where only links in the interior of the $k$-th time-slice are updated and spatial links on the time-slice boundaries are fixed. 
This requires locality, which is a property of the standard Wilson plaquette action. 
Wilson loop averages $\langle W_{r \times T} \rangle$ are obtained by computing the average of the right hand side of Eq.\ (\ref{eq:multilevel_Wilson}), 
which contains the time-slice averages $[\mathds{P}_k]$. Polyakov loop correlation functions can be computed in almost the same way, just replacing Eq.\ (\ref{eq:multilevel_Wilson}) by
\begin{align}
\label{eq:multilevel_Polyakov} L^*(x) L(x+r \hat{j}) = \{[\mathds{P}_1] [\mathds{P}_2] \dots [\mathds{P}_{N_\textrm{ts}}]\}_{\alpha \alpha \beta \beta} .
\end{align}

When inserting chromoelectric fields using the cloverleaf discretization (\ref{EQN_cloverleaf}), 
as done in the case of Wilson loops, one has to choose their positions in such a way that they do not contain interior links of two time-slices. 
Clearly, this requires times-lices of thickness $p_j \geq 2$. Moreover, for Wilson loops the spatial Wilson lines $\mathds{L}$ should be located on time-slice boundaries. 
This implies certain restrictions for the temporal extent of the Wilson loop $T$ for given time-slice partitioning. For example with time-slice partitioning $p_\text{ts} = \{ 2 \}$, 
only Wilson loops with temporal extent $T/a = 2, 6, 10, \ldots$ can be computed, since the chromoelectric field is always inserted at the center of one of the two temporal Wilson lines. 
For efficiency reasons, one should select a time-slice partitioning, which allows one to compute a large number of different temporal extents $T$. 
The time-slice partitionings we used for the Wilson loops are collected in Table~\ref{TAB002}. 
For Polyakov loops such restrictions do not exist and we chose the simple partitioning $p_\textrm{ts} = \{ T/10 a \}$ with chromoelectric fields always inserted at $t = a$.

\begin{table}[htb]
\begin{center}

\def\arraystretch{1.2}
\begin{tabular}{ccccc}
\hline
Ensemble & Simulation & $p_\text{ts}$ & $r/a$ & $T/a$ \\
\hline
A & 1 & $\{ 1,1,1,1,1,1,1,1,2^\ast \}$ & $\{2,3,\dots,10\}$ & $\{5,6,\dots,18\}$ \\ 
B & 2 & $\{ 1,1,1,1,1,1,1,1,2^\ast \}$ & $\{2,3,\dots,13\}$ & $\{2,3,\dots,18\}$ \\
C & 3 & $\{ 1,1,1,1,1,1,1,1,2^\ast \}$ & $\{2,3,\dots,8\}$ & $\{2,3,\dots,17\}$ \\
C & 4 & $\{ 1,1,1,1,1,1,1,1,2^\ast \}$ & $\{9,10,\dots,15\}$ & $\{2,3,\dots,17\}$ \\
C & 5 & $\{ 1,1,1,1,1,1,2^\ast,1,1,1,1,1,$ & $\{2,3,\dots,15\}$ & $\{18,19\}$ \\
  &   & ${}\quad\ \ \, 1,2^\ast,1,1,1,1,1,1,2^\ast,1,1,1,1,$ & & $\{20,21,22,23\}$ \\
  &   & ${}\quad\ \ \,  1,1,2^\ast,1,1,1,1,1,1,2^\ast,1,1,1,$ & & \\
  &   & ${} \: 1,1,1,2,1,1,1,1,1,1,2,4\}$ & & \\
\hline
\end{tabular}

\end{center}
\caption{\label{TAB002}Time-slice partitionings for the computation of Wilson loops with the multilevel algorithm. Chromoelectric fields are inserted at time-slices marked with $^\ast$.}
\end{table}

A multilevel simulation includes the following steps:
\begin{itemize}
\item[(0)] Start with any gauge link configuration (we use ``hot starts'' for our Wilson loop simulations, i.e.\ randomly chosen gauge links, and ``cold starts'' for our Polyakov loop simulations, 
i.e.\ all gauge links set to unity). Perform $n_{u,\text{th}}$ heatbath sweeps to generate a thermalized gauge link configuration, where each sweep is followed by $n_\text{or}$ overrelaxation steps.

\item[(1)] Generate $n_m$ sublattice configurations for each of the $n_\text{ts}$ time-slices by updating the links in the interior $n_m n_u$ times using the a heatbath algorithm 
(the $n_m$ sublattice configurations are then separated by $n_u$ updates).

\item[(2)] Compute time-slice averages $[\mathds{P}_k]$ on the corresponding $n_m$ sublattice configurations.

\item[(3)] Compute Wilson loop or Polyakov loop samples on the full gauge link configuration according to Eq.\ (\ref{eq:multilevel_Wilson}) or Eq.\ (\ref{eq:multilevel_Polyakov}), 
respectively, using the time-slice averages $[\mathds{P}_k]$ obtained in step~(2).

\item[(4)] Generate the next full gauge link configuration by performing $n_{u,0}$ heatbath sweeps, where each sweep is followed by $n_\text{or}$ overrelaxation steps.

\item[(5)] Repeat steps (1) to (4) $n_{m,0}$ times and estimate $\langle W_{r \times T} \rangle$ or $\langle L^*(0) L(r) \rangle$, respectively, from the samples obtained in step~(3).
\end{itemize}
Gauge link configurations for our Wilson loop computations and our Polyakov loop computations were generated with different simulation codes, 
with the CL2QCD software package \cite{Philipsen:2014mra} and a code developed in Ref.\ \cite{Banerjee:2011ra}. The values of the simulation parameters, 
which were crudely optimized by numerical tests (see Appendix~\ref{sec:appendix}), are collected in Table \ref{TAB008}.

\begin{table}[htb]
\begin{center}

\def\arraystretch{1.2}
\begin{tabular}{cccccccc}
\hline
Ensemble & Loops & $n_{m,0}$ & $n_{u,\text{th}}$ & $n_{u,0}$ & $n_\text{or}$ & $n_m$ & $n_u$ \\
\hline
A & Wilson   & $\phantom{0}800$ & $\phantom{0}1000$ & $\phantom{0}20$ & $\phantom{0}1$ & $\phantom{00}50$ & $2$ \\
  & Polyakov & $1284$           & $\phantom{00}150$ & $\phantom{00}5$ & $\phantom{0}3$ & $6000$           & $1$ \\
B & Wilson   & $\phantom{0}800$ & $40000$           & $200$           & $15$           & $\phantom{00}50$ & $2$ \\
  & Polyakov & $1825$           & $\phantom{00}150$ & $\phantom{00}5$ & $\phantom{0}3$ & $3000$           & $1$ \\
C & Wilson   & $\phantom{0}800$ & $40000$           & $200$           & $15$           & $\phantom{00}50$ & $2$ \\
  & Polyakov & $1391$           & $\phantom{00}150$ & $\phantom{00}5$ & $\phantom{0}3$ & $6000$           & $1$ \\
\hline
\end{tabular}
\end{center}
\caption{\label{TAB008}Multilevel simulation parameters.}
\end{table}



\section{\label{sec:result}Numerical results}

The majority of quantities discussed in the following are lattice quantities and, thus, depend on the lattice spacing $a$. 
In contrast to section~\ref{SEC_lattice}, we suppress the $a$-dependence throughout this section, to keep the notation simple. 
For example $V(r,a)$ from section~\ref{SEC_lattice} is equivalent to $V(r)$ in this section. 
Moreover, we exclusively use tree-level improved separations $r_I$, as discussed in section~\ref{SEC_lattice}. 
For simplicity we omit the index, i.e.\ denote separations just by $r$.


\subsection{The static potential}

To have a reference, we first computed the static potential $V(r)$ using Eq.\ (\ref{EQN_V_lat}). 
For each value $r = a, 2a, 3a, \ldots$ we fitted a constant to $V_\textrm{eff}(r,T)$ in the range $T_\textrm{min} \leq T \leq T_\textrm{max}$.
$T_\textrm{min}$ and $T_\textrm{max}$ were chosen sufficiently large, to guarantee a strong suppression of excited states, using an algorithm discussed in section~5 of Ref.\ \cite{Capitani:2018rox}:
\begin{itemize}
\item $T'_\textrm{min}$ is the minimal $T$, where $V_\textrm{eff}(r,T)$ and $V_\textrm{eff}(r,T+a)$ differ by less than $2 \, \sigma$.

\item $T'_\textrm{max}$ is the maximum $T$, where correlation functions have been computed (see Table~\ref{TAB002}).

\item Fit constants $V(r)$ to $V_\textrm{eff}(r,T)$ for all ranges $T_\textrm{min} \ldots T_\textrm{max}$ with $T'_\textrm{min} \leq T_\textrm{min}$, $T_\textrm{max} \leq T'_\textrm{max}$ 
and $T_\textrm{max} - T_\textrm{min} \geq 2 \, a$. Results with $\chi^2_\text{red} > 1.0$ are discarded, 
where $\chi^2_\text{red}$ denotes the uncorrelated reduced $\chi^2$ of the corresponding fit. If all fits yield $\chi^2_\text{red} > 1.0$, 
keep that one with the smallest $\chi^2_\text{red}$ and discard all others.

\item As the final result for $V(r)$ take the fit result corresponding to the longest plateau, i.e.\ with
maximum $T_\textrm{max} - T_\textrm{min}$. If there are several fit results with the same maximum $T_\textrm{max} - T_\textrm{min}$, 
take the fit result with the smallest $T_\textrm{min}$.
\end{itemize}
The fit results represent the static potential $V$ at tree-level improved separations according to Appendix~B of Ref.\ \cite{Necco:2001xg}. 
We find agreement with results from the literature \cite{Guagnelli:1998ud,Necco:2001xg}. 
Similarly, we computed the static potential at finite, but small temperature $1 / T = 1/N_t a \approx 1 / 4.8 \, r_0 \approx 82 \, \textrm{MeV}$ from Polyakov loops,
%
%
\begin{eqnarray}
\label{EQN_V_LL} V(r) = -\frac{1}{T} \ln \langle L^\dag(0) L(r)\rangle
\end{eqnarray}
with the Polyakov loops separated along one of the spatial coordinate axes, e.g.\ $L(r) = L(\mathbf{r} = (0,0,r))$. $V(r)$ obtained via Eq.\ (\ref{EQN_V_LL}) 
should be almost identical to $V(r)$ obtained via Eq.\ (\ref{EQN_V_lat}), 
because the leading finite temperature correction in $\langle L^\dag(0) L(r)\rangle$ is suppressed by $e^{-(E_1(r) - V(r)) T} \approx e^{-(3/r_0) \times 4.8 r_0} \approx 5 \times 10^{-7}$, 
as can be read off from Eq.\ (\ref{EQN_LL_spectral_decomposition}) (the crude estimate $E_1(r) - V(r) \approx 3/r_0$ was taken from Ref.\ \cite{Capitani:2018rox}). 
This is supported by our numerical results as well as by numerical results from Refs.\ \cite{Kaczmarek:2003dp,Bala:2019cqu}. 
For spatial separations $r/a \gtapprox 3 a$ and $r \gtapprox 0.13 \, \text{fm}$ the lattice results for $V(r)$ can be parametrized by the Cornell potential
\begin{eqnarray}
\label{EQN635} V_\textrm{Cornell}(r) = V_0 - \frac{\alpha}{r} + \sigma r
\end{eqnarray}
(see e.g.\ Ref.\ \cite{Karbstein:2018mzo}). Performing a fit to the lattice results for $V(r)$ obtained with Wilson loops and ensemble~B in the range $2.889 \leq r/a \leq 14.012$ 
we find $\alpha = 0.260$ and $\sigma = 1.508 / r_0^2$, in reasonable agreement with results from the literature, e.g.\ Ref.\ \cite{Koma:2007jq}. 
The additive constant $V_0$ is divergent in the continuum limit and physically irrelevant.
%


\subsection{Numerical proof of concept}

Now we consider $F_E(r) / F_E(r^\ast)$, where $F_E$ is the non-renormalized force defined in Eq.\ (\ref{eq:f1}), when using Wilson loops. 
We determined $F_E(r)$ by fitting a constant to $F_{E,\text{eff}}(r)$, where the $T$ range for the fit was chosen in the same way as for the static potential (see the discussion at the beginning of section~\ref{sec:result}). 
When using Polyakov loops, we determine $F_E(r)$ by Eq.\ (\ref{EQN634}). 
In both cases $r^\ast = 0.48 \, r_0 \approx 0.24 \, \textrm{fm}$ is a fixed separation chosen such that $r^\ast / a$ is an integer for all three ensembles, 
i.e.\ $r^\ast/a = 4 , 5 , 6$ for ensembles A, B and C, respectively. 
Since we compute the static force at improved separations, we take the two data points $F_E(r_1)$ and $F_E(r_2)$ with $r_1$ and $r_2$ closest to and enclosing $r^\ast$ 
and interpolate with $\alpha/r^2 + \sigma$ to read off $F_E(r^\ast)$. 
Note that, because of the multiplicative renormalization of $F_E$ discussed in section~\ref{SEC_lattice}, $F_E(r) / F_E(r^\ast) = F_E^\text{ren}(r) / F_E^\text{ren}(r^\ast)$. 
Thus, also $F_E(r) / F_E(r^\ast)$ should exhibit only mild differences to $F(r)/F(t^\ast)$.

\begin{figure}[htb]
\begin{center}
\includegraphics[width=0.7\textwidth]{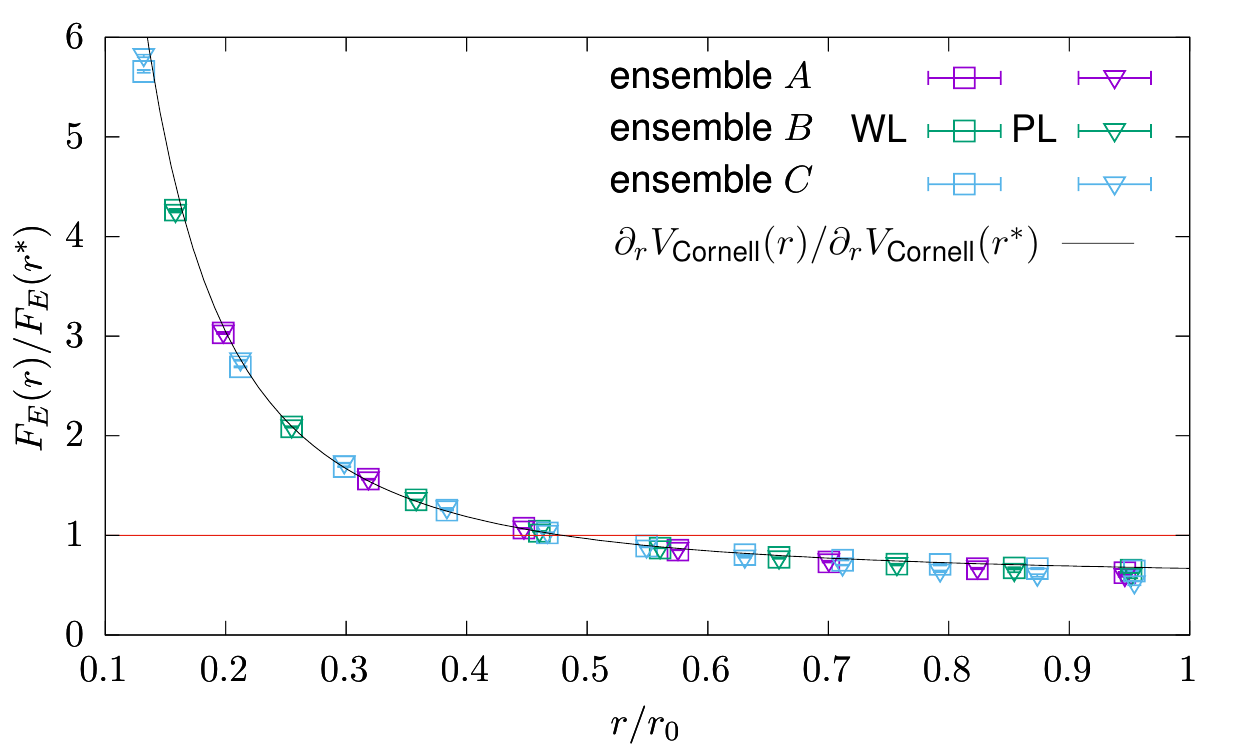}
\end{center}
\caption{\label{FIG001}$F_E(r) / F_E(r^\ast)$ as a function of $r$ for $r^\ast = 0.48 \, r_0 \approx 0.24 \, \textrm{fm}$ obtained from Wilson loops (boxes) and Polyakov loops (triangles). 
For comparison we also show $\partial_r V_\textrm{Cornell}(r) / \partial_r V_\textrm{Cornell}(r^\ast)$.}
\end{figure}

In Fig.~\ref{FIG001} we show $F_E(r) / F_E(r^\ast)$ as a function of the separation for all three ensembles obtained from Wilson loops as well as from Polyakov loops. 
For comparison we also show \\ $\partial_r V_\textrm{Cornell}(r) / \partial_r V_\textrm{Cornell}(r^\ast)$, 
which represents the same physical quantity, this time, 
however, obtained from the lattice result for the static potential parametrized according to Eq.\ (\ref{EQN635}) and not from a direct computation of the static force. 
The agreement of $F_E(r) / F_E(r^\ast)$ and $\partial_r V_\textrm{Cornell}(r) / \partial_r V_\textrm{Cornell}(r^\ast)$ is a numerical proof of concept for our method of computing the static force.

\subsection{\label{SEC644}The renormalization factor $Z_E$}

\begin{figure}[htb]
\begin{center}
\includegraphics[width=0.49\textwidth,page=1]{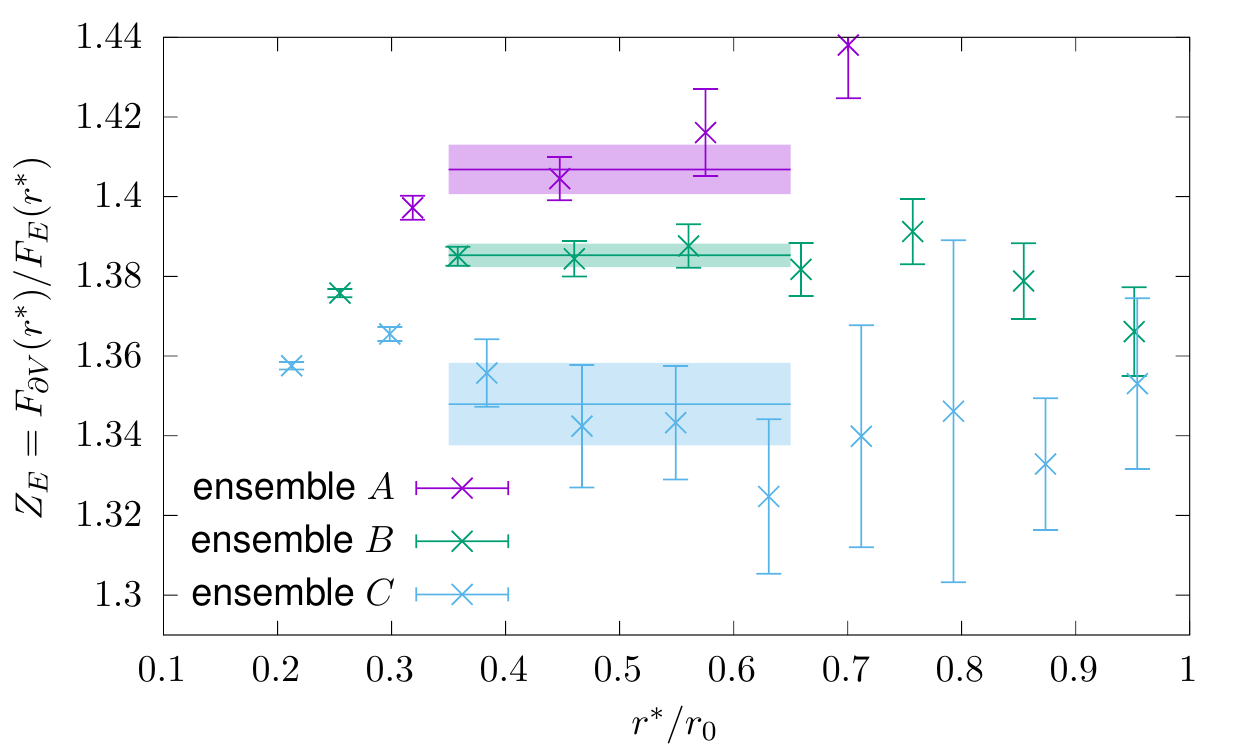}
\includegraphics[width=0.49\textwidth,page=2]{Z_E.pdf}
\end{center}
\caption{\label{FIG002}$Z_E = F_{\partial V}(r^\ast) / F_E(r^\ast)$ as a function of $r^\ast$. 
The colored horizontal lines and error bands represent the fits to determine a numerical value for $Z_E$ for each ensemble. \textbf{(left)}~Wilson loops. \textbf{(right)}~Polyakov loops.}
\end{figure}

In Fig.~\ref{FIG002} we show the renormalization factor $Z_E = F_{\partial V}(r^\ast) / F_E(r^\ast)$, defined in Eq.\ (\ref{EQN_ZE}), 
as a function of $r^\ast$, both for Wilson loops (left plot) and for Polyakov loops (right plot). 
As discussed in section~\ref{SEC_lattice}, $Z_E$ should be fairly independent of $r^\ast$. 
This expected constant behavior of $Z_E$ is confirmed by our numerical results, which exhibit clear plateaus. 
There is, however, a dependence on $\beta$ and, thus, on the lattice spacing $a$, where $Z_E$ is slowly decreasing for decreasing $a$. 
This is consistent with our expectation $Z_E(a) \rightarrow 1$ for $a \rightarrow 0$ discussed in section~\ref{SEC_lattice}.

We determined the numerical values for $Z_E$ separately for Wilson loops and for Polyakov loops 
and for each of our three ensembles by fitting a constant to the lattice data points shown in Fig.~\ref{FIG002} in the range $0.35 \, r_0 \leq r^\ast \leq 0.65 \, r_0$. 
The fit results are collected in Table~\ref{TAB007}.

\begin{table}[htb]
\begin{center}
\def\arraystretch{1.2}
\begin{tabular}{cccc}
\hline
Ensemble & $a$ $(\text{fm})$ & $Z_E$ from Wilson loops & $Z_E$ from Polyakov loops \\
\hline
A & $0.060$ & $1.4068(63)$ & $1.4001(20)$ \\
B & $0.048$ & $1.3853(30)$ & $1.3776(10)$ \\
C & $0.040$ & $1.348(11)\phantom{0}$ & $1.3628(13)$ \\
\hline
\end{tabular}
\end{center}
\caption{\label{TAB007}Renormalization factors $Z_E$ obtained by fitting constants to $F_{\partial V}(r^\ast) / F_E(r^\ast)$ in the range $0.35 \, r_0 \leq r^\ast \leq 0.65 \, r_0$.}
\end{table}


\subsection{Comparison of efficiency: $F_{\partial V}$ versus $F_E$}


\subsubsection{\label{SEC385}Asymptotic $T$ behavior of Wilson loops}

The spectral decomposition of Wilson loops and of Wilson loops with chromoelectric field insertions has been discussed in section~\ref{sec:FE} and section~\ref{sec:theory:analytic}.
For large $T$, the Wilson loop $\langle \textrm{Tr}\{{\rm P} \, W_{r \times T}\} \rangle$ is proportional to $e^{-V(r) T}$ with leading order correction suppressed by $e^{-(E_1(r) - V(r)) T}$ (see Eq.\ (\ref{spcdecW})). 
In contrast, the Wilson loop with chromoelectric field insertion $\langle \textrm{Tr}\{{\rm P} \, W_{r \times T} g E_j\}\rangle$ 
has a leading order correction proportional to $e^{-(E_1(r) - V(r)) T / 2}$ (see Eq.\ (\ref{spcdecWE_})), i.e.\ a correction more weakly suppressed with respect to the temporal separation $T$. 
Thus, to determine the static force using $F_E$, we expect that one has to consider correlation functions at $T$ values around twice as large compared to using $F_{\partial V}$, 
to get a similar suppression of unwanted contributions by excited states.

\begin{figure}[htb]
\begin{center}
\includegraphics[width=0.49\textwidth]{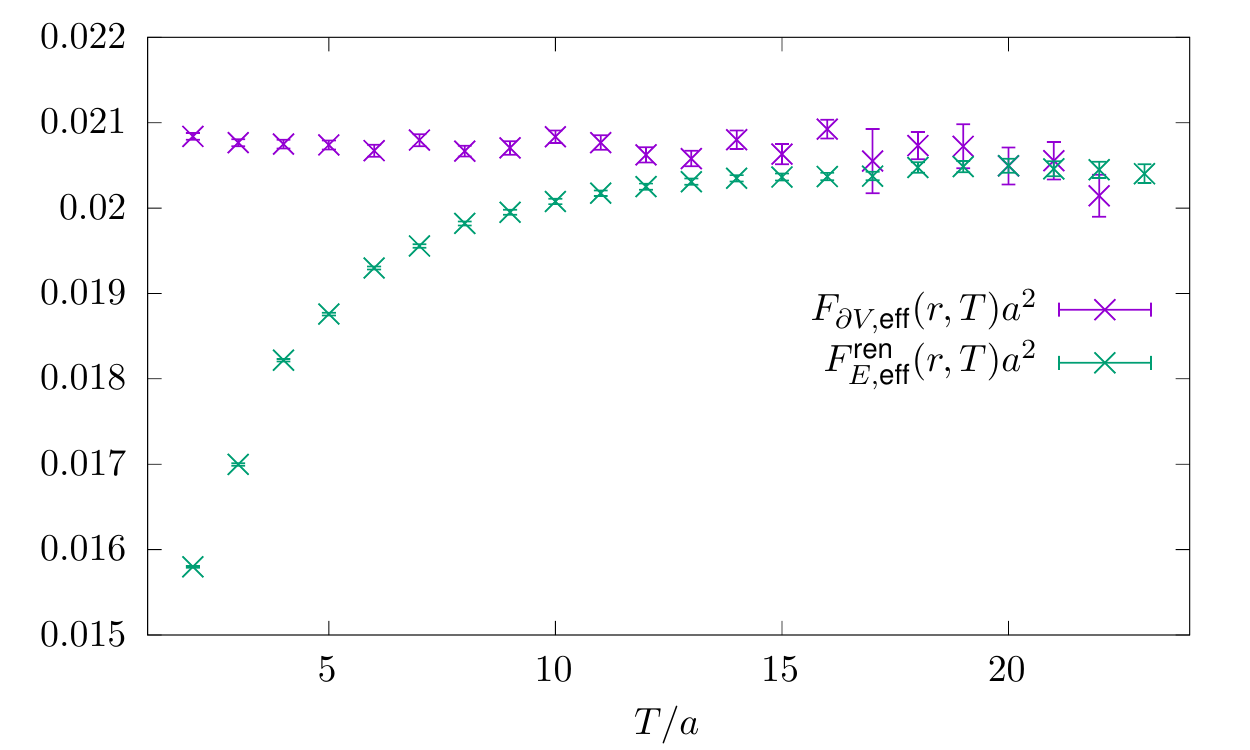}
\end{center}
\caption{\label{FIG005}Asymptotic $T$ behavior of $F_{E,\text{eff}}^\text{ren}(r,T)$ and of $F_{\partial V,\text{eff}}(r,T)$ for $r/a = 5$ and ensemble~C.}
\end{figure}

In Fig.~\ref{FIG005} we compare the asymptotic $T$ behavior of the renormalized effective force
\begin{eqnarray}
F_{E,\text{eff}}^\text{ren}(r,T) = Z_E F_{E,\text{eff}}(r,T)
\end{eqnarray}
with $Z_E$ taken from Table~\ref{TAB007} and of
\begin{eqnarray}
F_{\partial V,\text{eff}}(r,T) = \frac{V_\text{eff}(r+a,T) - V_\text{eff}(r-a,T)}{2 a}
\end{eqnarray}
for spatial separation $r/a = 5$ and ensemble~C. $F_{E,\text{eff}}^\text{ren}(r,T)$ converges to a plateau only at $T/a \gtapprox 14$. $F_{\partial V,\text{eff}}(r,T)$, 
on the other hand, is essentially constant already for $T/a \gtapprox 6$, which is consistent with our theoretical expectation.

For Polyakov loops there is no such difference between the two methods. In both cases contributions by excited states are proportional to $e^{-(E_1(r) - V(r)) T}$, where $T$ denotes the temporal extent of the lattice.


\subsubsection{Statistical precision and computing time}

Now we compare the efficiency of different approaches to compute the static force. A useful quantity to assess the efficiency is $(\Delta O)^2 \tau$, 
the computing time $\tau$ needed to compute an observable $O$ with statistical error $\Delta O$. Small values indicate an efficient method, 
large values an inefficient method. Since statistical errors in Monte Carlo simulations are proportional to $1 / \sqrt{\tau}$, 
this quantity allows a simple and fair comparison of two methods, even if the times invested for the corresponding computations are different.

\begin{figure}[htb]
\begin{center}
\includegraphics[width=0.49\textwidth,page=2]{ML.pdf}
\includegraphics[width=0.49\textwidth,page=1]{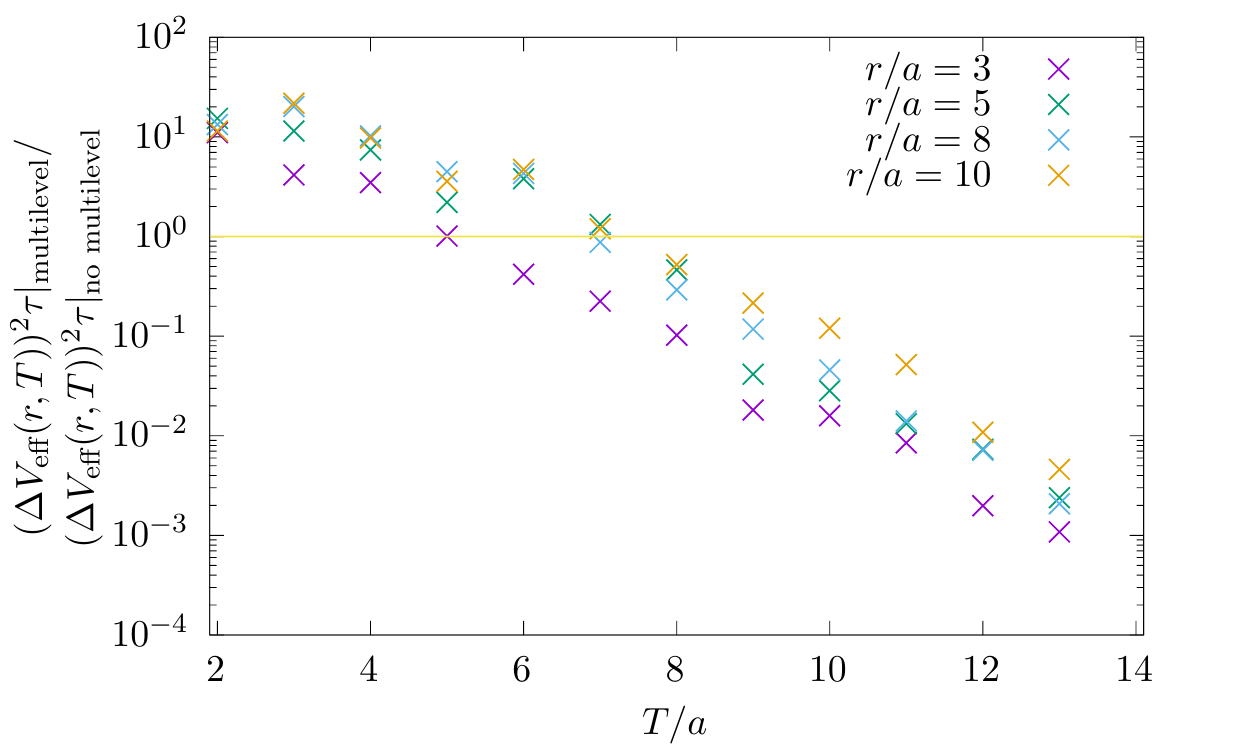}
\end{center}
\caption{\label{FIG003}Comparison of the efficiency, when using the multilevel algorithm and when not using the multilevel algorithm, for ensemble~C. \textbf{(left)}~Computation of the effective force $F_{E,\text{eff}}(r,T)$. \textbf{(right)}~Computation of the effective potential $V_\text{eff}(r,T)$.}
\end{figure}

We start by exploring the benefit of using the multilevel algorithm for the case of Wilson loops. 
To this end we compare in Fig.~\ref{FIG003} a computation, where we employed the multilevel algorithm with optimized parameters, 
as discussed in Appendix~\ref{sec:appendix}, to a standard heatbath simulation without multilevel algorithm using the ratio $(\Delta O)^2 \tau|_\text{multilevel} / (\Delta O)^2 \tau|_\text{no multilevel}$. 
The observable investigated in the left plot is the effective force, i.e.\ $O = F_{E,\text{eff}}(r,T)$, 
for several fixed separations $r$. The observable investigated in the right plot is the effective potential, i.e.\ $O = V_\text{eff}(r,T)$, for the same fixed separations $r$. 
Both plots show that at small temporal separations $T$ it is even more expensive to use the multilevel algorithm. 
However, at large temporal separations, which are typically needed for a precise extraction of the static force or the static potential, there is a huge gain in efficiency, when employing the multilevel algorithm. 
For example for ensemble C and $T/a = 14$ the time needed to compute the static force with the same statistical precision is reduced by a factor of around $10^4$ almost independent of $r$. 
As one can read off from Fig.~\ref{FIG005}, it is necessary to compute the correlation functions at such temporal separations.

\begin{figure}[htb]
\begin{center}
\includegraphics[width=0.49\textwidth,page=1]{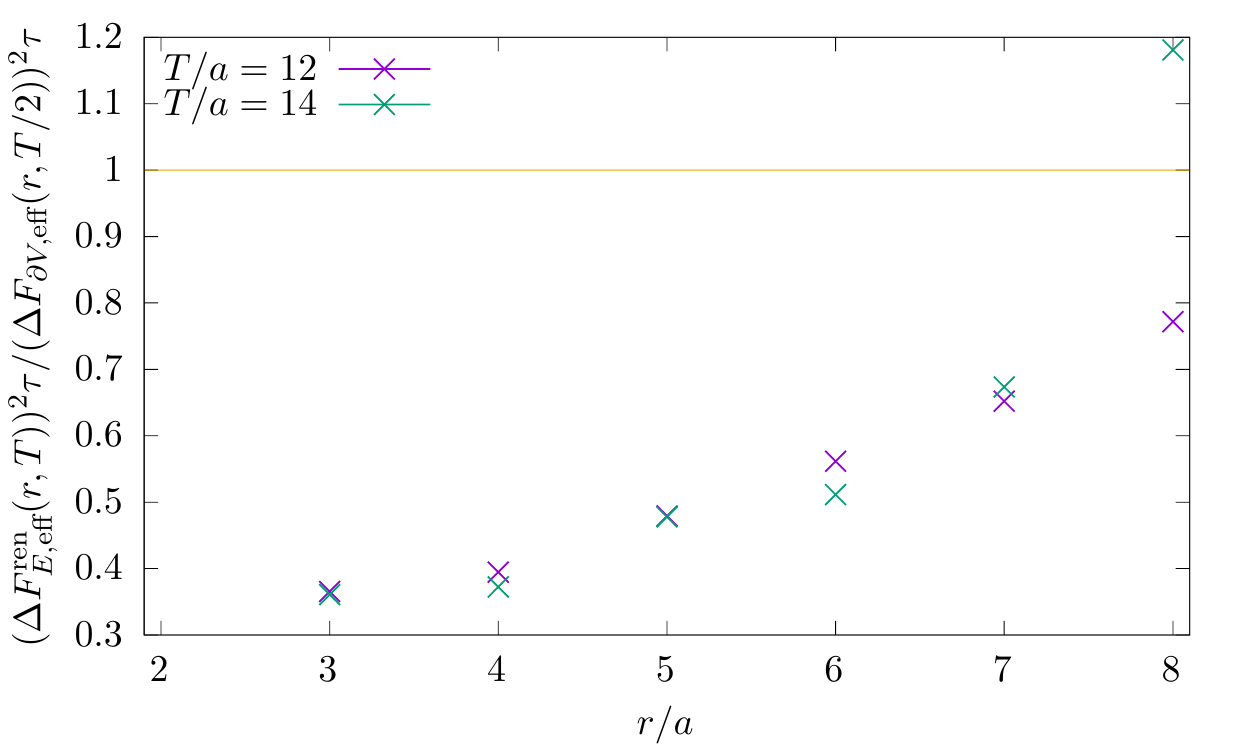}
\includegraphics[width=0.49\textwidth,page=2]{drV_vs_F.pdf}
\end{center}
\caption{\label{FIG004}Comparison of the efficiency, when computing the static force using $F_E$ and when using $F_{\partial V}$, for ensemble~C. \textbf{(left)}~Wilson loops. \textbf{(right)}~Polyakov loops.}
\end{figure}

In Fig.~\ref{FIG004} we compare the efficiency of computing the static force, when using $F_E$ and when using $F_{\partial V}$. 
Again we show ratios of the quantity $(\Delta O)^2 \tau$. 
The left plot shows the result for Wilson loops, where $O = F_{E,\text{eff}}^\text{ren}(r,T)$ 
in the numerator and $O = F_{\partial V,\text{eff}}(r,T/2)$ in the denominator 
(note that roughly twice as large temporal separations $T$ are needed for $F_{E,\text{eff}}^\text{ren}$ compared to $F_{\partial V,\text{eff}}$ for a similar suppression of contributions by excited states; see section~\ref{SEC385}). 
The right plot shows the result for Polyakov loops, where $O = F_E^\text{ren}(r)$ in the numerator and $O = F_{\partial V}(r)$ in the denominator. 
The plots indicate that it is advantageous to compute the static force via $F_E$, when using Wilson loops and typical temporal and spatial separations, 
while for Polyakov loops the traditional method via $F_{\partial V}$ is significantly more efficient. 
This different behavior is somewhat surprising and should be investigated in more detail in the future.

\section{\label{sec:conclusion}Conclusions}

We tested a novel method to compute the static force $F(r)$ from expectation values of Wilson or Polyakov loops with chromoelectric field insertions, 
which was suggested in Refs.\ \cite{Vairo:2015vgb,Vairo:2016pxb}. 
The numerical results converge well towards the continuum limit, after a multiplicative renormalization by
$Z_E$, where $Z_E=1+\mathcal{O}(g_0^2)$ is computed non-perturbatively from the standard force 
$F_{\partial V}$.
Concerning efficiency, our method appears to be comparable to the traditional method of first computing the static potential and then taking the derivative. 
For Wilson loops we even found a slight advantage, while for Polyakov loops our method seems to be somewhat less efficient. 
In this exploratory study we used pure SU(3) gauge theory, lattices with rather small spatial volume and the multilevel algorithm. 
In the future it will be interesting to study the applicability and efficiency of our method on full QCD gauge link ensembles with larger spatial volume, where the multilevel algorithm is not available.

This explorative computation of the static force also constitutes an important preparatory step for future projects, 
where similar correlation functions (Wilson or Polyakov loops with chromoelectric or chromomagnetic field insertions) need to be computed. 
An example is the computation of $1/m$ and $1/m^2$ corrections ($m$ denotes the heavy quark mass) to the ordinary static potential, 
as obtained in potential Non-Relativistic QCD in Refs.\ \cite{Brambilla:2000gk,Pineda:2000sz,Brambilla:2004jw} (see also Refs.\ \cite{Eichten:1979pu,Barchielli:1988zp}) 
and evaluated on the lattice in Refs.\ \cite{Bali:1997am,Bali:2000gf,Koma:2006si,Koma:2006fw}, 
or to hybrid static potentials, as theoretically worked out and suggested in Refs.\ \cite{Oncala:2017hop,Brambilla:2018pyn,Brambilla:2019jfi}.



\section*{Acknowledgements}

N.B., V.L.\ and A.V.\ thank S.\ Datta for providing a simulation code used in Ref.\ \cite{Banerjee:2011ra}. N.B.\ thanks Gunnar Bali for several discussions.

C.R.\ acknowledges support by a Karin and Carlo Giersch Scholarship of the Giersch foundation. M.W.\ acknowledges funding by the Heisenberg Programme of the Deutsche Forschungsgemeinschaft (DFG, German Research Foundation) -- Projektnummer 399217702. This work has been supported by the NSFC and the Deutsche Forschungsgemeinschaft (DFG, German Research Foundation) through the funds provided to the Sino-German Collaborative Research Center TRR110 ``Symmetries and the Emergence of Structure in QCD'' (NSFC Grant No.\ 12070131001, DFG Project-ID 196253076 -- TRR 110), and by the DFG cluster of excellence ``ORIGINS''.

Calculations on the Goethe-HLR and on the FUCHS-CSC high-performance computer of the Frankfurt University were conducted for this research. We would like to thank HPC-Hessen, funded by the State Ministry of Higher Education, Research and the Arts, for programming advice. Part of the simulations have been carried out on the computing facilities of the Computational Center for Particle and Astrophysics (C2PAP) of the cluster of excellence ORIGINS.



\appendix

\section{\label{sec:appendix}Optimization of multilevel parameters}

\begin{figure}[htb]
\begin{center}
\includegraphics[width=0.49\textwidth,page=1]{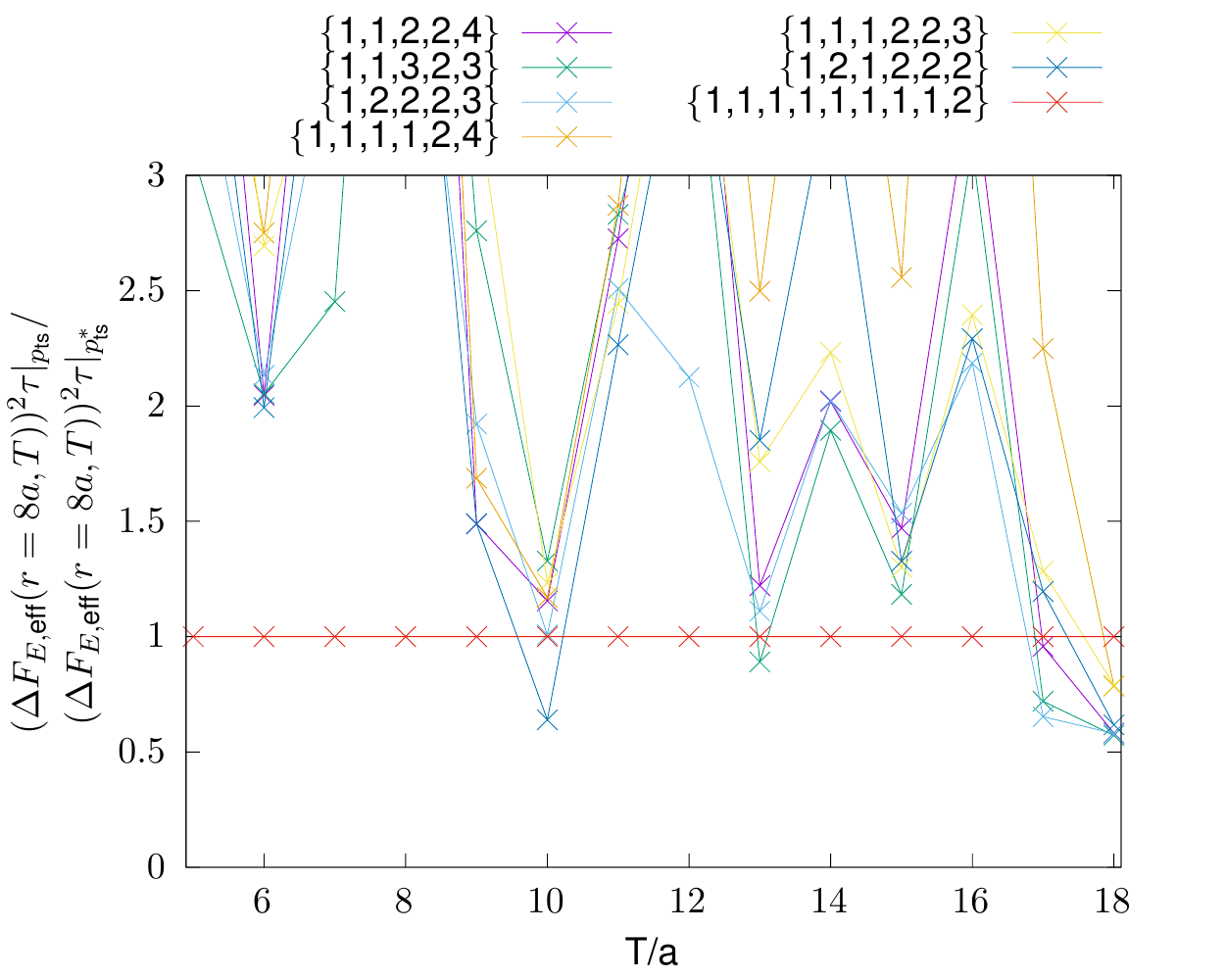} \\
\includegraphics[width=0.49\textwidth,page=2]{mlvl_param.pdf}
\includegraphics[width=0.49\textwidth,page=3]{mlvl_param.pdf}
\end{center}
\caption{\label{FIG006}Dependence of the efficiency of the multilevel algorithm on various numerical parameters, when computing $F_{E,\text{eff}}(r = 8a,T)$ on ensemble~A. \textbf{(top)}~$p_\text{ts}$ compared to the optimal $p_\text{ts}^\ast = \{ 1,1,1,1,1,1,1,1,2 \}$ for $n_m = 100$, $n_u = 1$. \textbf{(bottom left)}~$n_u$ compared to the optimal $n_u^\ast = 2$ for $n_m = 100$, $p_\text{ts} = \{ 1,1,1,1,1,1,1,1,2 \}$. \textbf{(bottom right)}~$n_m$ compared to the optimal $n_m^\ast = 50$ for $n_u = 2$, $p_\text{ts} = \{ 1,1,1,1,1,1,1,1,2 \}$.}
\end{figure}

To crudely optimize the multilevel parameters for our Wilson loop computations on ensemble~A, 
we first compared several time-slice partitionings $p_\text{ts}$ for fixed $n_m = 100$ and $n_u = 1$. Fig.~\ref{FIG006} (top plot) indicates that thinner time-slices are more efficient. 
Thus, we decided for time-slice partitionings $p_\text{ts}$, where the majority of time-slices have thickness $p_j = 1$ (see Table~\ref{TAB002}).

In a second step we investigated, how efficiency is related to the parameter $n_u$, the number of heatbath updates between two successive sublattice configurations. 
For fixed $n_m = 100$ we find $n_u = 2$ as optimum (see Fig.~\ref{FIG006}, bottom left plot).

Finally, we investigated, how efficiency is related to the parameter $n_m$, the number of sublattice configurations used to compute time-slice averages $[\mathds{P}_k]$. 
For fixed $n_u = 2$ we identify $n_m = 50$ as a rather efficient choice (see Fig.~\ref{FIG006}, bottom right plot).

Similar optimizations were carried out for ensembles B and C and for the Polyakov loop computations. 

\begin{thebibliography}{99}

\bibitem{Wilson:1974sk}
K.~G.~Wilson,
``Confinement of quarks,''
Phys.\ Rev.\ D \textbf{10}, 2445-2459 (1974).

\bibitem{Susskind:1976pi}
L.~Susskind,
``Coarse grained quantum chromodynamics,''
in {\it Les Houches 1976, Proceedings, Weak and Electromagnetic Interactions At High Energies}, Amsterdam 1977, 207.

\bibitem{Fischler:1977yf}
W.~Fischler,
``Quark-anti-quark potential in QCD,''
Nucl.\ Phys.\ B \textbf{129}, 157-174 (1977).

\bibitem{Brown:1979ya}
L.~S.~Brown and W.~I.~Weisberger,
``Remarks on the static potential in quantum chromodynamics,''
Phys.\ Rev.\ D \textbf{20}, 3239 (1979).

\bibitem{Anzai:2009tm}
C.~Anzai, Y.~Kiyo and Y.~Sumino,
``Static QCD potential at three-loop order,''
Phys.\ Rev.\ Lett.\ \textbf{104}, 112003 (2010)
[arXiv:0911.4335 [hep-ph]].

\bibitem{Smirnov:2009fh}
A.~V.~Smirnov, V.~A.~Smirnov and M.~Steinhauser,
``Three-loop static potential,''
Phys.\ Rev.\ Lett.\ \textbf{104}, 112002 (2010)
[arXiv:0911.4742 [hep-ph]].

\bibitem{Brambilla:1999qa}
N.~Brambilla, A.~Pineda, J.~Soto and A.~Vairo,
``The Infrared behavior of the static potential in perturbative QCD,''
Phys.\ Rev.\ D \textbf{60}, 091502 (1999)
[arXiv:hep-ph/9903355 [hep-ph]].

\bibitem{Brambilla:1999xf}
N.~Brambilla, A.~Pineda, J.~Soto and A.~Vairo,
``Potential NRQCD: an effective theory for heavy quarkonium,''
Nucl.\ Phys.\ B \textbf{566}, 275 (2000)
[arXiv:hep-ph/9907240 [hep-ph]].

\bibitem{Pineda:2000gza}
A.~Pineda and J.~Soto,
``The Renormalization group improvement of the QCD static potentials,''
Phys.\ Lett.\ B \textbf{495}, 323-328 (2000)
[arXiv:hep-ph/0007197 [hep-ph]].

\bibitem{Brambilla:2006wp}
N.~Brambilla, X.~Garcia i Tormo, J.~Soto and A.~Vairo,
``The Logarithmic contribution to the QCD static energy at N$^4$LO,''
Phys.\ Lett.\ B \textbf{647}, 185-193 (2007)
[arXiv:hep-ph/0610143 [hep-ph]].

\bibitem{Brambilla:2009bi}
N.~Brambilla, A.~Vairo, X.~Garcia i Tormo and J.~Soto,
``The QCD static energy at NNNLL,''
Phys.\ Rev.\ D \textbf{80}, 034016 (2009)
[arXiv:0906.1390 [hep-ph]].

\bibitem{Aoki:2019cca}
S.~Aoki \textit{et al.} [Flavour Lattice Averaging Group],
``FLAG Review 2019: Flavour Lattice Averaging Group (FLAG),''
Eur.\ Phys.\ J.\ C \textbf{80}, no.\ 2, 113 (2020)
[arXiv:1902.08191 [hep-lat]].

\bibitem{Karbstein:2014bsa}
F.~Karbstein, A.~Peters and M.~Wagner,
``${\Lambda}_{\overline{\mathrm{MS}}}^{(n_f=2)}$ from a momentum space analysis of the quark-antiquark static potential,''
JHEP \textbf{09}, 114 (2014)
[arXiv:1407.7503 [hep-ph]].

\bibitem{Bazavov:2014soa}
A.~Bazavov, N.~Brambilla, X.~Garcia i Tormo, P.~Petreczky, J.~Soto and A.~Vairo,
``Determination of $\alpha_s$ from the QCD static energy: an update,''
Phys.\ Rev.\ D \textbf{90}, no.\ 7, 074038 (2014)
[erratum: Phys.\ Rev.\ D \textbf{101}, no.\ 11, 119902 (2020)]
[arXiv:1407.8437 [hep-ph]].

\bibitem{Karbstein:2018mzo}
F.~Karbstein, M.~Wagner and M.~Weber,
``Determination of $\Lambda_{\overline{\textrm{MS}}}^{(n_f=2)}$ and analytic parametrization of the static quark-antiquark potential,''
Phys.\ Rev.\ D \textbf{98}, no.\ 11, 114506 (2018)
[arXiv:1804.10909 [hep-ph]].

\bibitem{Takaura:2018vcy}
H.~Takaura, T.~Kaneko, Y.~Kiyo and Y.~Sumino,
``Determination of $\alpha_s$ from static QCD potential: OPE with renormalon subtraction and lattice QCD,''
JHEP \textbf{04}, 155 (2019)
[arXiv:1808.01643 [hep-ph]].

\bibitem{Bazavov:2019qoo}
A.~Bazavov \textit{et al.} [TUMQCD],
``Determination of the QCD coupling from the static energy and the free energy,''
Phys.\ Rev.\ D \textbf{100}, no.\ 11, 114511 (2019)
[arXiv:1907.11747 [hep-lat]].

\bibitem{Ayala:2020odx}
C.~Ayala, X.~Lobregat and A.~Pineda,
``Determination of $\alpha(M_z)$ from an hyperasymptotic approximation to the energy of a static quark-antiquark pair,''
JHEP \textbf{09}, 016 (2020)
[arXiv:2005.12301 [hep-ph]].

\bibitem{Pineda:1998id}
A.~Pineda,
``Heavy quarkonium and nonrelativistic effective field theories,''
Ph.D.\ thesis, University of Barcelona (1998).

\bibitem{Hoang:1998nz}
A.~H.~Hoang, M.~C.~Smith, T.~Stelzer and S.~Willenbrock,
``Quarkonia and the pole mass,''
Phys.\ Rev.\ D \textbf{59}, 114014 (1999)
[arXiv:hep-ph/9804227 [hep-ph]].

\bibitem{Necco:2001xg}
S.~Necco and R.~Sommer,
``The $N(f) = 0$ heavy quark potential from short to intermediate distances,''
Nucl.\ Phys.\ B \textbf{622}, 328-346 (2002)
[arXiv:hep-lat/0108008 [hep-lat]].

\bibitem{Necco:2001gh}
S.~Necco and R.~Sommer,
``Testing perturbation theory on the $N(f) = 0$ static quark potential,''
Phys.\ Lett.\ B \textbf{523}, 135-142 (2001)
[arXiv:hep-ph/0109093 [hep-ph]].

\bibitem{Vairo:2015vgb}
A.~Vairo,
``A low-energy determination of $\alpha_s$ at three loops,''
EPJ Web Conf.\ \textbf{126}, 02031 (2016)
[arXiv:1512.07571 [hep-ph]].

\bibitem{Vairo:2016pxb}
A.~Vairo,
``Strong coupling from the QCD static energy,''
Mod.\ Phys.\ Lett.\ A \textbf{31}, no.\ 34, 1630039 (2016).

\bibitem{Brambilla:2000gk}
N.~Brambilla, A.~Pineda, J.~Soto and A.~Vairo,
``The QCD potential at $O(1/m)$,''
Phys.\ Rev.\ D \textbf{63}, 014023 (2000)
[arXiv:hep-ph/0002250 [hep-ph]].

\bibitem{Baker:2018mhw}
M.~Baker, P.~Cea, V.~Chelnokov, L.~Cosmai, F.~Cuteri and A.~Papa,
``Isolating the confining color field in the SU(3) flux tube,''
Eur.\ Phys.\ J.\ C \textbf{79}, no.\ 6, 478 (2019)
[arXiv:1810.07133 [hep-lat]].

\bibitem{Baker:2019gsi}
M.~Baker, P.~Cea, V.~Chelnokov, L.~Cosmai, F.~Cuteri and A.~Papa,
``The confining color field in SU(3) gauge theory,''
Eur.\ Phys.\ J.\ C \textbf{80}, no.\ 6, 514 (2020)
[arXiv:1912.04739 [hep-lat]].

\bibitem{Luscher:2001up}
M.~L\"uscher and P.~Weisz,
``Locality and exponential error reduction in numerical lattice gauge theory,''
JHEP \textbf{09}, 010 (2001)
[arXiv:hep-lat/0108014 [hep-lat]].

\bibitem{Brambilla:2019zqc}
N.~Brambilla \textit{et al.} [TUMQCD],
``Static force from the lattice,''
PoS \textbf{LATTICE2019}, 109 (2019)
[arXiv:1911.03290 [hep-lat]].

\bibitem{Jahn:2004qr}
O.~Jahn and O.~Philipsen,
``The Polyakov loop and its relation to static quark potentials and free energies,''
Phys.\ Rev.\ D \textbf{70}, 074504 (2004)
[arXiv:hep-lat/0407042 [hep-lat]].

\bibitem{Eichten:1979pu}
E.~Eichten and F.~L.~Feinberg,
``Spin dependent forces in heavy quark systems,''
Phys.\ Rev.\ Lett.\ \textbf{43}, 1205 (1979).


\bibitem{Gattringer:2010zz}
C.~Gattringer and C.~B.~Lang,
``Quantum chromodynamics on the lattice,''
Lect.\ Notes Phys.\ \textbf{788}, 1-343 (2010).

\bibitem{Sommer:1993ce}
R.~Sommer,
``A New way to set the energy scale in lattice gauge theories and its applications to the static force and $\alpha_{\rm s}$ in SU(2) Yang--Mills theory,''
Nucl.\ Phys.\ B \textbf{411}, 839-854 (1994)
[arXiv:hep-lat/9310022 [hep-lat]].

\bibitem{Bali:1997am}
G.~S.~Bali, K.~Schilling and A.~Wachter,
``Complete $O(v^2)$ corrections to the static interquark potential from SU(3) gauge theory,''
Phys.\ Rev.\ D \textbf{56}, 2566-2589 (1997)
[arXiv:hep-lat/9703019 [hep-lat]].

\bibitem{Huntley:1986de}
A.~Huntley and C.~Michael,
``Spin-spin and spin-orbit potentials from lattice gauge theory,''
Nucl.\ Phys.\ B \textbf{286}, 211-230 (1987).

\bibitem{Bali:2000gf}
G.~S.~Bali,
``QCD forces and heavy quark bound states,''
Phys.\ Rept.\ \textbf{343}, 1-136 (2001)
[arXiv:hep-ph/0001312 [hep-ph]].

\bibitem{Lepage:1992xa}
G.~P.~Lepage and P.~B.~Mackenzie,
``On the viability of lattice perturbation theory,''
Phys.\ Rev.\ D \textbf{48}, 2250-2264 (1993)
[arXiv:hep-lat/9209022 [hep-lat]].

\bibitem{Koma:2006fw}
Y.~Koma and M.~Koma,
``Spin-dependent potentials from lattice QCD,''
Nucl.\ Phys.\ B \textbf{769}, 79-107 (2007)
[arXiv:hep-lat/0609078 [hep-lat]].

\bibitem{Christensen:2016wdo}
C.~Christensen and M.~Laine,
``Perturbative renormalization of the electric field correlator,''
Phys.\ Lett.\ B \textbf{755}, 316-323 (2016)
[arXiv:1601.01573 [hep-lat]].

\bibitem{Guazzini:2007bu}
D.~Guazzini \textit{et al.} [ALPHA],
``Non-perturbative renormalization of the chromo-magnetic operator in heavy quark effective theory and the $B^\ast$-$B$ mass splitting,''
JHEP \textbf{10}, 081 (2007)
[arXiv:0705.1809 [hep-lat]].

\bibitem{APE:1987ehd}
M.~Albanese \textit{et al.} [APE],
``Glueball Masses and String Tension in Lattice QCD,''
Phys. Lett. B \textbf{192}, 163-169 (1987)

\bibitem{Jansen:2008si}
K.~Jansen \textit{et al.} [ETM],
``The Static-light meson spectrum from twisted mass lattice QCD,''
JHEP \textbf{12}, 058 (2008)
[arXiv:0810.1843 [hep-lat]].

\bibitem{Philipsen:2014mra}
O.~Philipsen, C.~Pinke, A.~Sciarra and M.~Bach,
``CL$^2$QCD -- Lattice QCD based on OpenCL,''
PoS \textbf{LATTICE2014}, 038 (2014)
[arXiv:1411.5219 [hep-lat]].

\bibitem{Banerjee:2011ra}
D.~Banerjee, S.~Datta, R.~Gavai and P.~Majumdar,
``Heavy quark momentum diffusion coefficient from lattice QCD,''
Phys.\ Rev.\ D \textbf{85}, 014510 (2012)
[arXiv:1109.5738 [hep-lat]].

\bibitem{Capitani:2018rox}
S.~Capitani, O.~Philipsen, C.~Reisinger, C.~Riehl and M.~Wagner,
``Precision computation of hybrid static potentials in SU(3) lattice gauge theory,''
Phys.\ Rev.\ D \textbf{99}, no.\ 3, 034502 (2019)
[arXiv:1811.11046 [hep-lat]].

\bibitem{Guagnelli:1998ud}
M.~Guagnelli \textit{et al.} [ALPHA],
``Precision computation of a low-energy reference scale in quenched lattice QCD,''
Nucl.\ Phys.\ B \textbf{535}, 389-402 (1998)
[arXiv:hep-lat/9806005 [hep-lat]].

\bibitem{Kaczmarek:2003dp}
O.~Kaczmarek, F.~Karsch, P.~Petreczky and F.~Zantow,
``Heavy quark free energies, potentials and the renormalized Polyakov loop,''
Nucl.\ Phys.\ B Proc.\ Suppl.\ \textbf{129}, 560-562 (2004)
[arXiv:hep-lat/0309121 [hep-lat]].

\bibitem{Bala:2019cqu}
D.~Bala and S.~Datta,
``Nonperturbative potential for the study of quarkonia in QGP,''
Phys.\ Rev.\ D \textbf{101}, no.\ 3, 034507 (2020)
[arXiv:1909.10548 [hep-lat]].

\bibitem{Koma:2007jq}
Y.~Koma, M.~Koma and H.~Wittig,
``Relativistic corrections to the static potential at $O(1/m)$ and $O(1/m^2)$,''
PoS \textbf{LATTICE2007}, 111 (2007)
[arXiv:0711.2322 [hep-lat]].

\bibitem{Pineda:2000sz}
A.~Pineda and A.~Vairo,
``The QCD potential at $O(1/m^2)$: complete spin dependent and spin independent result,''
Phys.\ Rev.\ D \textbf{63}, 054007 (2001)
[erratum: Phys.\ Rev.\ D \textbf{64}, 039902 (2001)]
[arXiv:hep-ph/0009145 [hep-ph]].

\bibitem{Brambilla:2004jw}
N.~Brambilla, A.~Pineda, J.~Soto and A.~Vairo,
``Effective field theories for heavy quarkonium,''
Rev.\ Mod.\ Phys.\ \textbf{77}, 1423 (2005)
[arXiv:hep-ph/0410047 [hep-ph]].

\bibitem{Barchielli:1988zp}
A.~Barchielli, N.~Brambilla and G.~M.~Prosperi,
``Relativistic corrections to the quark-anti-quark potential and the quarkonium spectrum,''
Nuovo Cim.\ A \textbf{103}, 59 (1990).

\bibitem{Koma:2006si}
Y.~Koma, M.~Koma and H.~Wittig,
``Nonperturbative determination of the QCD potential at $O(1/m)$,''
Phys.\ Rev.\ Lett.\ \textbf{97}, 122003 (2006)
[arXiv:hep-lat/0607009 [hep-lat]].

\bibitem{Oncala:2017hop}
R.~Oncala and J.~Soto,
``Heavy quarkonium hybrids: spectrum, decay and mixing,''
Phys.\ Rev.\ D \textbf{96}, no.\ 1, 014004 (2017)
[arXiv:1702.03900 [hep-ph]].

\bibitem{Brambilla:2018pyn}
N.~Brambilla, W.~K.~Lai, J.~Segovia, J.~Tarr\'us Castell\`a and A.~Vairo,
``Spin structure of heavy-quark hybrids,''
Phys.\ Rev.\ D \textbf{99}, no.\ 1, 014017 (2019)
[erratum: Phys.\ Rev.\ D \textbf{101}, no.\ 9, 099902 (2020)]
[arXiv:1805.07713 [hep-ph]].

\bibitem{Brambilla:2019jfi}
N.~Brambilla, W.~K.~Lai, J.~Segovia and J.~Tarr\'us Castell\`a,
``QCD spin effects in the heavy hybrid potentials and spectra,''
Phys.\ Rev.\ D \textbf{101}, no.\ 5, 054040 (2020)
[arXiv:1908.11699 [hep-ph]].

\end{thebibliography}
\end{document}